# Higgs in a box: investigating the nature of a scientific discovery

Julia Woithe[1], Margherita Boselli[1], Panagiota Chatzidaki[1,2], Merten Nikolay Dahlkemper[1,3], Ruadh Duggan[1,4], Guillaume Durey[1], Niklas Herff[1,5], Anja Kranjc Horvat[1], Daniele Molaro[1], Gernot Werner Scheerer[1], Sascha Schmeling[1], Patrick Georges Thill[1], Jeff Wiener[1], Sarah Zoechling[1,6]

(1) European Organization for Nuclear Research CERN, Esplanade des Particules 1, 1217 Meyrin, Switzerland
(2) Lund University, Lund University Physics Education Research (LUPER) group, Box 118, 221 00 Lund, Sweden
(3) University of Göttingen, Friedrich-Hund-Platz 1, 37077 Göttingen, Germany
(4) Utrecht University, Freudenthal Institute, Princetonplein 5, 3584 CC Utrecht, The Netherlands
(5) Technische Universität Dresden, Institute of Nuclear and Particle Physics, 01069 Dresden, Germany
(6) University of Vienna, AECC Physics, Porzellangasse 4/2/2, 1090 Vienna, Austria

## Abstract

The discovery of the Higgs boson by the ATLAS and CMS collaborations in 2012 concluded the longest search for a particle in the history of particle physics and was based on the largest and most complex physics experiments ever conducted, involving thousands of scientists and engineers from around the world. It provided crucial evidence for a theory developed in the 1960s that describes the existence of the invisible Brout-Englert-Higgs field and the effects of this field on the mass of elementary particles. After the discovery, the work on the theoretical prediction was awarded the Nobel Prize in Physics 2013. This discovery provides a prime example of modern science in the making and a fantastic opportunity to discuss important aspects of Nature of Science (NoS) in the classroom. In this article, we draw connections between a) milestones in the discovery of the Higgs boson, b) important aspects of NoS, and c) hands-on activities with mystery boxes, which are an effective tool to enable students to experience elements of scientific discovery and explicitly reflect on NoS. We hope that this supports educators in bringing lively discussions about modern physics research into their classrooms.

# 1. Introduction

On 4 July 2022, the particle physics community was in a festive mood. There were a series of exciting talks, with Nobel prize winners and numerous journalists coming to CERN. Why? Because on this day, CERN celebrated the 10-year anniversary of the discovery of the "Higgs boson". Indeed, on 4 July 2012, CERN witnessed the official announcement of the discovery of a new particle, which matched the prediction of a theory developed in the 1960s. The announcement concluded the longest search for a particle in the history of particle physics and was based on the largest and most complex physics experiments ever conducted, involving thousands of scientists and engineers from around the world.

The discovery of the Higgs boson by the ATLAS and CMS collaborations provided crucial evidence supporting a theory developed in the 1960s that describes the existence of the invisible Brout-Englert-Higgs (BEH) field and the effects of this field on the mass of elementary particles. In 2013, the Nobel Prize in Physics was awarded to François Englert and Peter Higgs for their work on the theory of how particles acquire mass [1]. While the BEH theory is only accessible using graduate level maths and theoretical physics skills, the Higgs boson discovery was prominently featured in many news outlets and soon made it into pop culture (e.g. [2]). Recent secondary school physics textbooks also include this remarkable story (see, e.g., [3]).

Indeed, describing the stories of scientific discoveries can be a powerful tool to teach Nature of Science (NoS) [4]. In particular, modern physics topics such as quantum physics show great potential for discussing aspects of NoS [5]. But textbook writers often fail to harness this potential by giving too little attention to NoS or representing it in an unfavourable way [6]. For example, even though an explicit-reflective approach is essential in effectively teaching NoS [7], [8], there are often no prompts for explicit reflection on NoS [9], [10].

We argue that the captivating discovery of the Higgs boson provides a prime example of modern physics in the making, and a fantastic opportunity to reflect on and discuss important aspects of NoS. In this article, we draw connections between a) milestones in the discovery of the Higgs boson, b) important aspects of NoS, and c) hands-on activities with mystery boxes, which are an effective tool to enable students to experience elements of scientific discovery and explicitly reflect on NoS, see Figure 1.

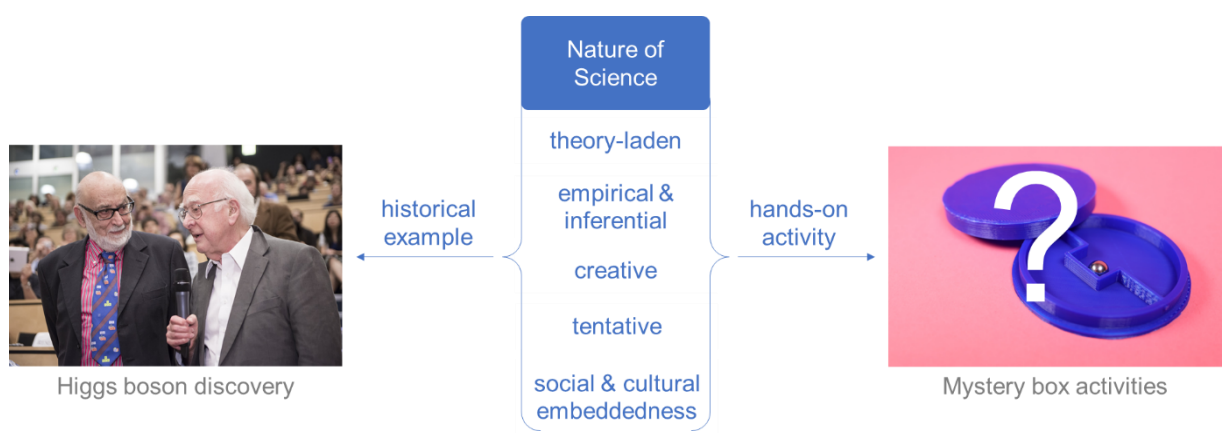

*Figure 1. Concept map showing links between the three different components of this paper.*

## 2. The discovery of the Higgs boson

This section highlights selected milestones leading to the discovery of the Higgs boson and is based on a book by Castillo [11]. We also suggest prompts for explicit reflection on NoS, which can accompany the text when using it in the classroom (see Box 1).

> **Box 1. Accompanying prompts for explicit reflection on NoS**
>
> Teaching NoS is most effective when using an explicit-reflective approach. Thus, we suggest accompanying this text about the Higgs boson discovery with the following prompts to guide students in identifying and reflecting on aspects of NoS.
>
> - How did theory guide the experiments?
> - Why is it impossible to observe Higgs bosons directly?
> - What indirect observations did the ATLAS collaboration make?
> - What did the ATLAS and CMS collaborations infer from their observations?
> - Which steps in the discovery process required creativity?
> - Which social dimensions of science were described?
> - Were scientists 100% certain that they discovered the Higgs boson?

**What was the initial problem?**

In the early 1960s, physicists already had a good theoretical understanding of the electromagnetic interaction, one of the four fundamental interactions that describes, for example, how electrically charged particles attract or repel each other. But they were struggling to find a similar theory for the other fundamental interactions. Specifically, there was one major problem related to the weak interaction, which describes, for example, how neutrons transform into protons in radioactive transformation processes ("beta minus decay"). The theory for the weak interaction only worked if the corresponding interaction particles, the so-called W- and Z-bosons, had zero mass. But from experiments, scientists knew that this was not the case.

**What was the idea to solve this problem?**

In 1964, three independent teams of physicists came up with an idea to fix this problem: Peter Higgs from the University of Edinburgh [12], Robert Brout and François Englert from the Université Libre de Bruxelles [13], and Gerald Guralnik, Carl R. Hagen and Tom Kibble from Imperial College London [14]. Building on ideas developed by Yoichiro Nambu [15], Jeffrey Goldstone [16] and Philip W. Anderson [17], they creatively adapted a mechanism called "spontaneous symmetry breaking" that was originally developed to explain the phenomenon of superconductivity. In particular, they came up with a theoretical trick to apply a similar mechanism to particle physics to solve the problem of massless particles.

At the time, the three teams did not know of each other's work but published three separate scientific articles with different approaches to a similar idea: What if the universe were "filled" with a special invisible field? The idea was that elementary particles would obtain their mass when they interacted with this field. If a particle interacted strongly with the field, the particle's mass would be high. If a particle only interacted a little bit, the particle's mass would be small. And if a particle did not interact with the field at all, it would remain massless. In this theory, mass is not considered a property of a particle, but the result of the interaction between each particle and the field. Over the following decades, as our understanding of particle physics evolved, this field became a cornerstone of the Standard Model of Particle Physics [18]. Today, this field is generally referred to as the Brout-Englert-Higgs (BEH) field.

Indeed, it is an interesting theory, but how can scientists test a theory involving an invisible field in an experiment? In one of his scientific articles, Peter Higgs wrote that there must be a special type of particle associated with the BEH field [19]. If the BEH field is disturbed enough, this particle could be created. Thus, scientists would only need to find this particle to get evidence for the existence of the BEH field. Since Peter Higgs was the only one to mention this particle in his publication in 1964, the particle associated with the BEH field was soon referred to as the Higgs boson [20].

So far, this idea sounds quite simple and intriguing but actually, the particle physics community was quite sceptical at first. For example, one of Peter Higgs' scientific articles was rejected at first by the journal "Physics Letters" because the editors thought it had no obvious relevance to physics. Therefore, to get his idea published, Peter Higgs had to rework his article which had to pass the scientific review process of a different journal [21].

**What was needed to test the idea of the BEH field experimentally?**

In the following, we discuss three things that were needed to test the idea of the BEH field experimentally: a) a better theoretical understanding of the idea, b) a particle accelerator providing enough energy to create Higgs bosons, and c) particle detectors making observations of Higgs bosons possible.

   a) A better theoretical understanding of the idea

In the 1970s, the idea of the BEH field gained popularity, and scientists started studying the properties of the (then hypothetical) Higgs boson. However, at that time the idea of the BEH field and the associated Higgs boson were extremely difficult to test experimentally. One big problem was that there were no theoretical estimates of the mass of the Higgs boson. In general, the mass of a particle is directly related to the amount of energy needed to create it, see Box 2. This then determines the scale of the experiment that is needed to make observations possible. Today we know that the mass of the Higgs boson is very large (125 GeV/$c^2$). This means that a very high energy is needed to create Higgs bosons and for this reason the idea of the BEH field could not be tested experimentally by searching for Higgs bosons at the time.

In a famous paper published in 1976, John Ellis, Mary K. Gaillard and Dimitri V. Nanopoulos described in detail how Higgs bosons could be created and how they might be observed. However, the authors concluded with a rather discouraging statement: "We apologise to experimentalists for having no idea what is the mass of the Higgs boson […]. For these reasons we do not want to encourage big experimental searches for the Higgs boson" [22].

> **Box 2. Mass–energy equivalence E=m·c² and the energy unit electronvolt**
>
> Einstein's famous formula E=m·c² describes that energy (E) and mass (m) are equivalent. This means that energy always corresponds to an equivalent amount of mass, and mass always corresponds to an equivalent amount of energy. We only need the square of the speed of light ($c^2$) as a conversion factor to express any mass in the form of energy. This means that we can actually think of mass as a form of energy, just like kinetic energy is a form of energy. In particle collisions, Einstein's formula plays an essential role: when two particles collide at high energies, their energy can transform into mass, that is, new particles can be created. In particular, the total amount of collision energy can transform into the energy corresponding to the mass of new particles that are created. If the mass of a particle is very large, a very high amount of collision energy is needed to create it.
>
> Particle physicists often use units other than the SI units Joule and Kilogram to express energies or masses. In particle physics, energies are given in the unit electronvolt, short eV, which is the product of the elementary charge e and the unit of voltage V. Thus, 1 eV equals $1.6 \cdot 10^{-19}$ Joule. Moreover, particle masses are given in the unit $eV/c^2$. For example, the Higgs boson mass is given at 125 $GeV/c^2$. This unusual unit makes it much easier to know how much collision energy is needed to create a certain type of particle. For example, a collision energy of at least 125 giga-electronvolts is needed to create one Higgs boson.

b) A particle accelerator providing enough energy to create Higgs bosons

Creating Higgs bosons in experiments requires sufficient energy to disturb the BEH field. This energy can be provided by particle accelerators. The first experiments that tried to create and observe Higgs bosons turned out empty-handed. They concluded that the mass of Higgs bosons must be quite large because their accelerators did not provide enough energy to create Higgs bosons. Yet, these experiments contributed significantly to our understanding because they narrowed down the possible mass range of the Higgs boson.

There were high hopes that large particle accelerators like the Large Electron Positron (LEP) Collider that was used from 1989 to 2000 at CERN in Switzerland or the Tevatron that was used from 1983 to 2011 at Fermilab in the US would allow creating and detecting Higgs bosons. And indeed, shortly before the planned shutdown of LEP, one of the experiments at LEP reported a suspicious signal in their data at a mass of around 115 $GeV/c^2$ [23]. However, as this result was not statistically significant at 3.2 sigma (see Box 3 for more information) and was not confirmed by the other experiments at LEP, it had a high probability of being caused only by a statistical fluctuation. Therefore, physicists requested to extend the machine's operational lifetime to collect more data in the hope of getting a clearer signal closer to the 5-sigma discovery standard (see Box 3). After intense discussions within the particle physics community, LEP was nonetheless shut down in 2000 to make space for a new accelerator providing higher energies and thus a much better chance of creating Higgs bosons: the Large Hadron Collider (LHC) [24].

> **Box 3: The 5-sigma discovery standard in particle physics**
>
> Scientists use a statistical hypothesis-testing framework to determine which of two hypotheses is consistent with the data they collect in experiments. For example, hypothesis one (also called the null hypothesis) could be "The Higgs boson does not exist" and hypothesis two (also called the alternative hypothesis) could be "The Higgs boson does exist".
>
> After collecting data, scientists calculate the probability that they would see a signal that is at least as strong as the one their data shows, while assuming that the null hypothesis is correct. For example, particle physicists calculated the probability that their data would show a signal consistent with that of the Higgs boson even if the Higgs boson did not exist - purely because of statistical fluctuations. It is important to note that this probability is *not* the probability that the null model is true.
>
> If this probability is very small, the null hypothesis is rejected in favour of the alternative hypothesis. What does "very small" mean? It depends on the field of science. In many fields, a probability below 5% would be small enough to reject the null model. In particle physics however, a much stricter standard has been developed over the past 50 years. This standard is called "5 sigma". 5 sigma corresponds to a tiny probability of $3 \cdot 10^{-5}$ %, a chance of 1 in 3.5 million, that a signal in the data is only caused by statistical fluctuations [25]. In particle physics, journals now only accept articles claiming a discovery if the 5-sigma standard is met.

Eight years later, in 2008, the LHC was switched on at CERN. This is a 27 km long ring about 100 m underground accelerating protons in opposite directions in two separate beam pipes and colliding them at four points. It took over 30 years of work from the first discussions about the accelerator in 1977 to accelerating the first protons in 2008 [26]. The construction cost of the LHC was several billion Euros [27], making it one of the most expensive scientific research tools ever built. CERN is funded by the taxpayers of its member states. However, to make an ambitious project like the LHC possible, CERN's scientists also had to convince the governments of additional countries, like the US or Japan, to contribute.

    c) Particle detectors making observations of Higgs bosons possible

The hope with the LHC was that colliding protons at unprecedented energies would provide sufficient collision energy to disturb the BEH field enough and create Higgs bosons, see Box 2. But how would scientists know whether this had worked? There are several challenges when it comes to "observing" Higgs bosons.

One problem is that it is not possible to observe Higgs bosons directly, they are unobservable entities. As a particle with a high mass, Higgs bosons are not stable and have a very short lifetime, so they quickly transform into other, lighter particles. Today, we estimate the lifetime of Higgs bosons at $10^{-22}$ seconds. That is far too short a time for the particle to leave any kind of direct signal in a particle detector. Therefore, scientists had to find creative ways to collect indirect evidence for the existence of Higgs bosons and in turn the BEH field.

Thus, many teams of theorists studied in detail how Higgs bosons could transform into other particles and what kind of signals these secondary particles would leave in particle detectors. Based on this work on the theory, scientists determined the best strategies to search for Higgs bosons. As an example, Higgs bosons can transform into pairs of high-energetic photons. Photons are stable and can be measured using particle detectors (see Figure 2). So instead of looking directly for Higgs bosons, particle physicists can look for proton collisions at the LHC in which two photons are created. But to complicate things even further, proton collisions at the LHC produce many, many pairs of photons, and most of these are created through other processes that have nothing to do with Higgs bosons. Indeed, the likelihood that Higgs bosons are created in proton collisions at the LHC is very small. This means that a lot of data had to be taken and a very careful statistical analysis was needed to separate background events from the actual Higgs-related signal [28].

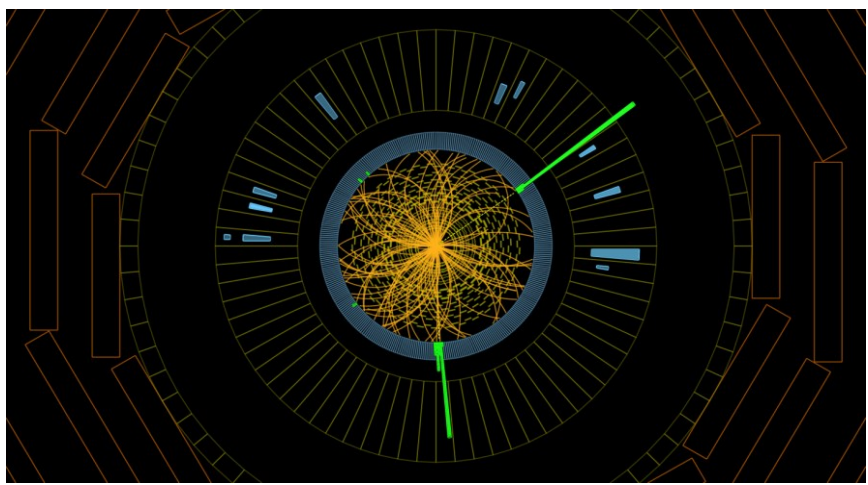

*Figure 2. Visualisation of a proton collision inside the CMS detector leading to a signal of two high-energetic photons (two long green lines). This could either show the signal of a Higgs boson that transformed into two photons or a background event. There is no way to know for sure what has caused a single event, but advanced statistical analyses based on huge amounts of data can help scientists draw conclusions about whether or not the Higgs boson exists. Reproduced from [29] © CERN (https://cern.ch/copyright).*

Two large collaborations consisting of thousands of physicists, engineers, technicians, students and support staff from hundreds of research institutes from around the world, the ATLAS and CMS collaborations, were formed to face the experimental challenge of collecting evidence for the existence of the Higgs boson. Both collaborations started separately to design, build, and operate complex detectors at two of the collision points of the LHC to record the signals produced by particle collisions.

Why two detectors? In addition to the 5-sigma standard (see Box 3), at least two independent observations are required to claim a discovery. Indeed, the detectors of the ATLAS and CMS collaborations used different detector components and strategies, and worked independently of each other to develop their analysis techniques.

Many additional challenges had to be overcome to realise indirect observations of Higgs bosons. For example, an impressive computing infrastructure was developed to handle, store, and analyse the data coming from the detectors.

**What was observed and how was this observation interpreted?**

In December 2011, the ATLAS and CMS collaborations presented the first hints of a new particle. However, the data did not yet allow physicists to draw any conclusions as the 5-sigma threshold was not crossed. Nevertheless, the particle physics community started getting excited. By July 2012, more data had been analysed and the CMS and ATLAS collaborations presented new results of their observations at CERN. Rumours had spread that Peter Higgs and François Englert were invited for these talks, two of the theorists that predicted the existence of the BEH field and Higgs boson almost 50 years before. As a result, people at CERN queued for hours, some in sleeping bags overnight, to get one of the seats in CERN's main auditorium.

On 4 July 2012, the spokespersons of the CMS and ATLAS collaborations, Joe Incandela and Fabiola Gianotti, presented the observations of the detectors and explained their interpretations. Figure 3 shows Fabiola Gianotti, who is today CERN's Director-General, during her presentation. When looking at pairs of high-energetic photons, ATLAS had observed an excess of a few hundred signals consistent with Higgs bosons, on top of a few thousand background events. Was this enough to claim a discovery? Not quite. However, when other transformation processes were also included and the data from ATLAS and CMS was combined, the results finally crossed the crucial 5 sigma threshold. Therefore, they inferred that their observations suggested a new particle with a mass around 125 GeV/$c^2$ [30]. They later published their combined results in an article that broke the record for the largest number of contributors to a single research article: 24 of the 33 pages of the article were needed to list all the 5154 authors and their institutions [31].

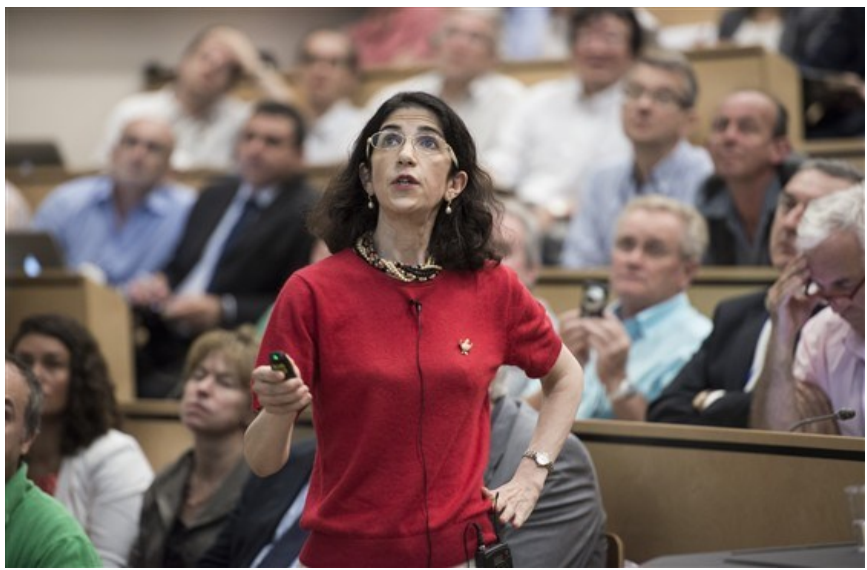

*Figure 3. Fabiola Gianotti presents the observations of the ATLAS collaboration and explains their interpretation on 4 July 2012 at CERN. Reproduced from* [32] © *CERN (https://cern.ch/copyright).*

But was this new particle the long sought-after Higgs boson? After the presentations of the ATLAS and CMS collaborations, CERN's Director-General at the time, Rolf Heuer, summarised "I think we have it", which was followed by an enthusiastic wave of applause and a tear of joy in Peter Higgs' eye. But it was not yet clear whether the particle observed was indeed the Higgs boson predicted by theory – or something else. Only in the following years more scientific evidence indicated that the newly discovered particle was indeed compatible with the Higgs boson. Finally, even a statistical significance of 5 sigma does not mean that we can be absolutely sure. In general, scientific knowledge is very reliable and durable, but never absolute or certain.

**What happened after the discovery?**

In October 2013, the Nobel Prize in Physics was awarded jointly to François Englert and Peter Higgs "for the theoretical discovery of a mechanism that contributes to our understanding of the origin of mass of subatomic particles, and which recently was confirmed through the discovery of the predicted fundamental particle, by the ATLAS and CMS experiments at CERN's Large Hadron Collider" [1].

Why only these two physicists? Sadly, Robert Brout, who co-authored a paper with François Englert [13], had died before the discovery was made. But what about all the other scientists who developed the theory, designed and built the LHC and the detectors, and analysed the data? Due to rules established more than 100 years ago, Nobel Prizes can only be awarded to a maximum of three people. Although the particle physics community celebrated the Nobel Prize, some started wondering whether these rules still make sense for today's research projects in which scientists do not work alone in a lab but instead contribute to the joint efforts of large research teams.

# 3. NoS and the nature of the Higgs boson discovery

Although science topics and methods are part of school curricula, many students have naive ideas about NoS [33]. A thorough understanding of NoS is crucial not just to understand recent scientific discoveries but also to tackle contemporary challenges of humankind. In the following, we provide a brief summary of important aspects of NoS relevant to school curricula, based on Abd-El-Khalick [34]. We connect these individual NoS aspects to the story of the discovery of the Higgs boson. Thereby, the following section refers to the prompts suggested in Box 1.

**Theory-laden NoS:** It is quite rare for scientists to start a scientific investigation with neutral observations of a phenomenon. Instead, scientists study scientific theories to determine research questions and to guide the investigation process (i.e. which observations to make (and which not to make), how to observe a phenomenon and how to interpret observations). Theories often postulate the existence of entities that cannot be observed directly. Instead, scientists collect indirect evidence to support and validate a theory. In particular, scientists derive new predictions from theory and they compare these predictions with their observations. Scientists' confidence in the tested theory increases if the observations match the initial predictions.

Connection to the Higgs boson discovery:

- The theory describing the BEH field predicted the existence of the Higgs boson. Both the BEH field and Higgs bosons cannot be observed directly.
- The theoretical understanding of the Higgs boson was used to calculate how it would transform into other particles that can be observed directly.
- Theory guided the strategies to indirectly collect evidence that supports the theory of the BEH field and the associated Higgs boson.

**Empirical & Inferential NoS:** Scientific claims depend on observations of natural phenomena. Here, observation can simply mean seeing something with one's own eyes or using other human senses. However, in most cases, scientists need to extend their senses by using measurement instruments because our eyes and other senses are not precise enough or simply not capable of measuring certain phenomena. In addition, if theories involve entities that cannot be observed directly, not even with very expensive measurement instruments, scientists can only gather indirect evidence. There is an important difference between an observation and how scientists interpret these observations based on scientific theories, that is, what they infer from the observation.

Connection to the Higgs boson discovery:

- The ATLAS and CMS collaborations built large particle detectors to make observations. Based on the recommendation of theorists, they observed, for example, pairs of high-energetic photons to gather indirect evidence for the existence of Higgs bosons.
- They compared the observation of the signal with the predicted background. After combining their research results, they inferred the existence of a new particle.

**Creative NoS:** Scientists often need to find creative solutions when doing scientific investigations. They might creatively adapt existing theories or use them in unusual contexts. They also need to find creative ways to do experiments, especially if they plan to study entities that are not directly observable. Moreover, creativity is needed when interpreting the data and generating conclusions.

Connection to the Higgs boson discovery:

- The theorists suggesting the BEH field creatively adapted a mechanism from a different field of physics to a problem in particle physics.
- Scientists and engineers developed innovative technologies to build the most powerful particle accelerator in the world.
- Physicists had to come up with clever ideas to gather evidence for the existence of Higgs bosons through indirect observations.

**Tentative NoS:** "Scientific knowledge is reliable and durable, but never absolute or certain." [34]. That means that scientific knowledge changes over time. For example, there may be new evidence due to technological advances or existing evidence may be interpreted in a new way due to advances in the development of scientific theories.

Connection to the Higgs boson discovery:

- Since the prediction of the Higgs boson in 1964, many theorists worked on improving our understanding of this particle and how it might show up in experiments.
- Technological advances made it possible to build a more powerful particle accelerator, the LHC, which provided enough energy to create Higgs bosons.
- One of the experiments at the Large Electron Positron (LEP) collider reported a suspicious signal in their data at a mass of around 115 GeV/c$^2$ but this turned out to be a statistical fluctuation and could not be confirmed by other experiments.
- CERN's Director-General in 2012, Rolf Heuer, said "I think we have it". At that time, it was still uncertain whether the newly discovered particle was indeed the predicted Higgs boson.

**Social dimensions of science & social and cultural embeddedness:** Individual scientists have subjective views on how to interpret data from an experiment. However, the social NoS helps to ensure its objectivity. The scientific communities have well-established procedures to critically examine and discuss new ideas. For example, scientists present their research at scientific conferences or submit research articles to journals to ensure that other scientists thoroughly review their ideas and findings. Moreover, science affects and is affected by the social and cultural environment. For example, it is often politicians who decide on funding for research projects.

Connection to the Higgs boson discovery:

- Peter Higgs submitted a scientific article to a journal where it was examined by anonymous reviewers. One of his articles was rejected so he had to rework his article to get his idea published in a different journal.
- To collect funding for the LHC, CERN's scientists had to convince politicians from many different governments to fund their research tools.
- The particle physics community developed the 5-sigma standard and requires observations at two independent experiments to claim a discovery.
- Awarding the Nobel Prize in Physics 2013 to only two theoretical physicists raised questions about whether the rules of the Nobel Prizes in science are outdated, as nowadays, scientific research is done by large teams working closely together.

# 4. Mystery boxes to teach NoS

In this section, we introduce mystery boxes as a teaching tool and explain how different mystery box activities can be used to represent important NoS aspects linked to the discovery of the Higgs boson. Even though the activities described below can be used as a stand-alone unit, we suggest having students first read and reflect on the story of the discovery of the Higgs boson before starting the activities. In this way, students can connect an example of a recent scientific discovery with their own scientific investigation while reflecting on aspects of NoS.

## 4.1. What are mystery boxes?

Mystery boxes (or black boxes) are objects with an internal structure that is invisible to an observer. However, observers can manipulate mystery boxes by changing different inputs and then observing the corresponding outputs [35]. Various mystery box activities exist. In their simplest form mystery boxes consist of a closed box (e.g. a shoebox) containing simple everyday or geometric items [36]. Two other commonly employed types are the mystery tube, which requires participants to pull on strings that are connected internally [37]; and the water-based mystery box, where one observes and compares inflows and outflows of water with a real device [38] or an online simulation [39].
All mystery boxes have the common feature that one can only indirectly observe their internal structure, which forces learners to conduct a series of systematic experiments to gather evidence about the contents of the mystery box. In a cyclic modelling process, students can usually quickly come up with a first model which they then test by deriving predictions and comparing these predictions to experimental observations.

Why use mystery boxes in the classroom? Mystery boxes are a versatile tool to teach NoS concepts in an explicit-reflective way [40], and they are effective across a wide range of ages ranging from kindergarten to secondary school [41]. They have been demonstrated to facilitate and foster scientific practices [42] and have been proven to significantly boost students' understanding of models and modelling [43].

## 4.2. A teaching sequence with mystery boxes

Below, we describe activities using cylindrical 3D-printable mystery boxes that contain a steel ball, see Figure 4 [44]. By moving the mystery box, a clicking sound indicates when the steel ball hits a surface and thus provides indirect observations of the inner structure. Furthermore, a small magnet can be used to steer the steel ball and systematically test previous models about the inner structure. However, the activities described can be adapted to other types of mystery boxes as well.

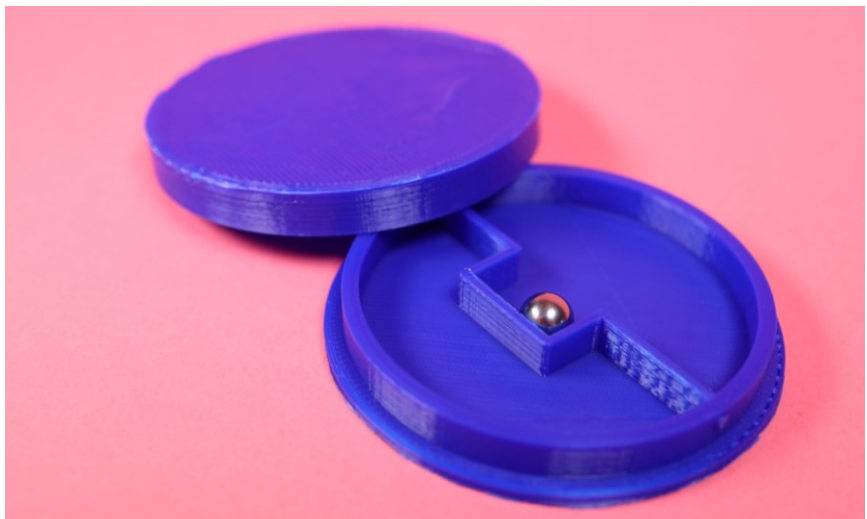

*Figure 4. Picture of the inner structure of a 3D-printable mystery box including a steel ball.*

**Preparation and general remarks**

For the following activities, we suggest that students work in teams of two to four to explore one mystery box together. Consequently, educators will need to prepare the following material for every team:
- One lid (see file "Lid.stl" [45])
- One base with the inner structure shown in Figure 5 (see file "Base_Heavyside_Step_Function.stl" [45])
- One 5 mm diameter steel ball (from ball bearings) or small sphere magnet
- One small Neodymium rod magnet

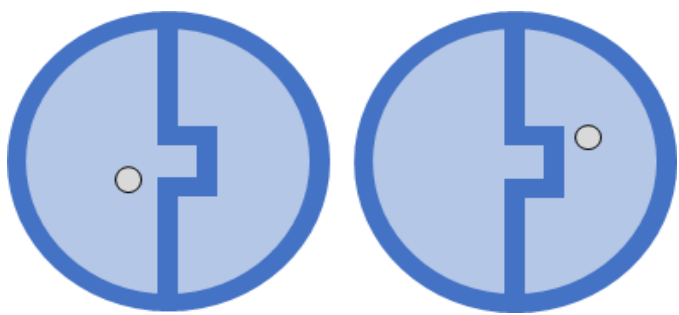

*Figure 5. Inner structure of the mystery box used in this series of activities. The grey circle represents a steel ball, which can be placed on either side of the inner wall. Depending on the level of the students, both options can be used to enrich the classroom discussion.*

The 3D-printable mystery boxes are designed with a bayonet lock closure. However, we strongly recommend gluing the lid on top of the base once the steel ball is inserted to prevent anyone from opening the box.

**Activity 1: The theory-laden NoS**

After a brief introduction, teachers demonstrate one of the mystery boxes (without the magnet at this stage) and show three drawings representing different models of their inner structure, see Figure 6. Based on these models, students come up with strategies to test them. Since it is not possible to see the inner structure directly, students develop ideas of how to make systematic observations using the acoustic signal of the steel ball moving inside the mystery box. For example, to test the idea of a triangular inner structure, students might suggest turning the box 360° and counting the number of "click" sounds.

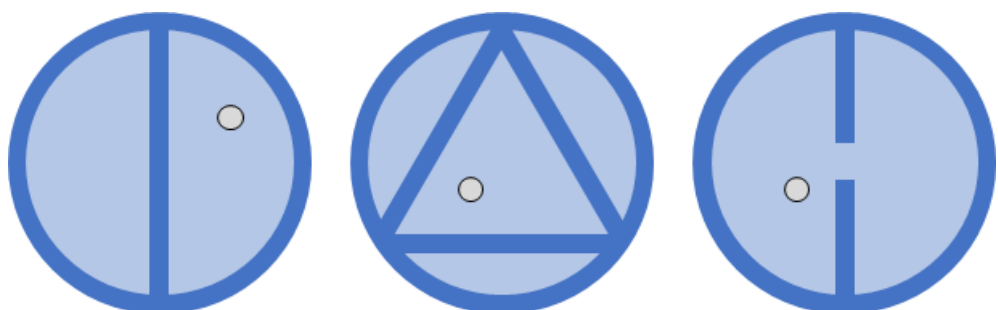

*Figure 6. Three different models of the possible inner structure of a cylindrical mystery box. The grey circle represents a steel ball.*

**Activity 2: The empirical & inferential NoS**

Now, teams of students receive one mystery box each to systematically test their ideas. From their indirect observations, they infer which of the three models best fits their observations. They note down their observations and the corresponding inferences, for example by using the poster template in Appendix 1.

**Activity 3: The creative NoS**

By design, none of the three initial models shown to the students (Figure 6) will perfectly fit the observations of the students. Only the teacher knows that the mystery boxes are based on the structure shown in Figure 5. Therefore, students are now encouraged to think creatively and come up with a new model that fits their observations better. If time permits, they can even design and build a new mystery box based on their new model and compare this new mystery box with the behaviour of the original mystery box. At the end of this activity, every student team draws a model representing their best idea of the inner structure of the mystery box, using the poster template in Appendix 1.

**Activity 4: The tentative NoS**

Now, students receive a small Neodymium rod magnet. With the help of this new research instrument, students test their model of the inner structure one more time by systematically moving the steel ball with the magnet and tracing out the inside wall. This advancement in their research method might lead to a better understanding of their mystery box. As a consequence, students might update their model one more time.

**Activity 5: Social dimensions of science & social and cultural embeddedness**

Students now present their results in front of the class, for example, in the form of a poster. First, they share their observations as well as their inferences when testing the three given models in activity 2. Then, they present their own model developed in activity 3 as well as the model that was improved based on the use of the magnet in activity 4. After all teams present their results, students are randomly assigned to give anonymous feedback to another group, for example by using post-it notes. In this way, every poster is reviewed by at least two students. This resembles the scientific peer-review process in which authors received anonymous feedback from independent reviewers. The authors of the posters can then use this feedback to improve their final model and their poster or explain why not to incorporate a reviewer's comment. Finally, all teams of students discuss the different models together with the goal to reach community consensus on the inner structure of their mystery box.

**Activity 6: Reflection on NoS**

After activities one to five, students explicitly reflect on their experience. In particular, they compare the different activities with the important milestones in the discovery of the Higgs boson. Here, the prompts in Box 1 can easily be adapted:

- How did theory guide your experiments?
- Why is it impossible to observe the inner structure of the mystery box directly?

- What indirect observations did you make?
- What did you infer from your observations?
- Which steps required creativity?
- Which social dimensions of science did you experience?
- Are you 100% certain that your final model is correct?

# 5. Conclusions

Many students have naive ideas about NoS [33]. Previous research recommends using mystery boxes [40] or the stories of (modern) scientific discoveries [4], [5] to teach NoS in an explicit-reflective way. In this article, we describe how to use the captivating story of the Higgs boson discovery together with a series of mystery box activities to reflect on important aspects of NoS. We believe that this will support teachers in enabling students not only to discover the story behind this prime example of a modern scientific discovery but also to deepen their understanding of NoS in general. Thus, students can develop the critical mindset and the scientific viewpoint that is needed to navigate in the world we live in.

# Acknowledgments

We would like to thank Matthew Chalmers for providing valuable feedback on the final draft of the manuscript.

# References

[1] Nobel Prize Outreach AB, 'The Nobel Prize in Physics 2013', *NobelPrize.org*, Oct. 08, 2013. https://www.nobelprize.org/prizes/physics/2013/summary/ (accessed Apr. 25, 2022).
[2] M. Cendrowski, C. Lorre, B. Prady, J. Reynolds, 'The Higgs Boson Observation', *The Big Bang Theory*, Chuck Lorre Productions; Warner Bros Television, Oct. 11, 2012.
[3] R. Sexl, H. Kühnelt, H. Stadler, P. Jakesch, and E. Sattlberger, *Sexl Physik 8*. Wien: öbv Österreichischer Bundesverlag Schulbuch GmbH & Co. KG, 2019. Accessed: Apr. 25, 2022. [Online]. Available: https://www.oebv.at/produkte/sexl-physik-8-schulbuch-0
[4] N. Nouri and W. F. McComas, 'History of science (HOS) as a vehicle to communicate aspects of nature of science (NOS): Multiple cases of HOS instructors' perspectives regarding NOS', *Res. Sci. Educ.*, vol. 51, no. 1, pp. 289–305, 2021, doi: https://doi.org/10.1007/s11165-019-09879-9.
[5] K. Stadermann and M. Goedhart, '"Why don't you just tell us what light really is?" Easy-to-implement teaching materials that link quantum physics to nature of science', *Phys. Educ.*, vol. 57, no. 2, p. 025014, Mar. 2022, doi: 10.1088/1361-6552/ac39e7.
[6] F. Abd-El-Khalick *et al.*, 'A longitudinal analysis of the extent and manner of representations of nature of science in U.S. high school biology and physics textbooks', *J. Res. Sci. Teach.*, vol. 54, no. 1, pp. 82–120, 2017, doi: 10.1002/tea.21339.
[7] F. Abd-El-Khalick and N. G. Lederman, 'The influence of history of science courses on students' views of nature of science', *J. Res. Sci. Teach. Off. J. Natl. Assoc. Res. Sci. Teach.*, vol. 37, no. 10, pp. 1057–1095, 2000, doi: https://doi.org/10.1002/1098-2736(200012)37:10<1057::AID-TEA3>3.0.CO;2-C.
[8] N. G. Lederman, 'Syntax of nature of science within inquiry and science instruction', in *Scientific inquiry and nature of science*, Springer, 2006, pp. 301–317.

[9] C. V. McDonald, 'Exploring representations of nature of science in Australian junior secondary school science textbooks', in *Representations of nature of science in school science textbooks*, 2017, pp. 98–117.

[10] W. Park, S. Yang, and J. Song, 'When Modern Physics Meets Nature of Science', *Sci. Educ.*, vol. 28, no. 9, pp. 1055–1083, Dec. 2019, doi: 10.1007/s11191-019-00075-9.

[11] L. Castillo, *The Search and Discovery of the Higgs Boson*. Institute of Physics, 2014. [Online]. Available: http://dx.doi.org/10.1088/978-1-6817-4078-2ch1

[12] P. W. Higgs, 'Broken symmetries, massless particles and gauge fields', *Phys. Lett.*, vol. 12, no. 2, pp. 132–133, Sep. 1964, doi: 10.1016/0031-9163(64)91136-9.

[13] F. Englert and R. Brout, 'Broken Symmetry and the Mass of Gauge Vector Mesons', *Phys. Rev. Lett.*, vol. 13, no. 9, pp. 321–323, Aug. 1964, doi: 10.1103/PhysRevLett.13.321.

[14] G. S. Guralnik, C. R. Hagen, and T. W. B. Kibble, 'Global Conservation Laws and Massless Particles', *Phys. Rev. Lett.*, vol. 13, no. 20, pp. 585–587, Nov. 1964, doi: 10.1103/PhysRevLett.13.585.

[15] Y. Nambu, 'Axial Vector Current Conservation in Weak Interactions', *Phys. Rev. Lett.*, vol. 4, no. 7, pp. 380–382, Apr. 1960, doi: 10.1103/PhysRevLett.4.380.

[16] J. Goldstone, 'Field theories with « Superconductor » solutions', *Il Nuovo Cimento*, vol. 19, no. 1, pp. 154–164, Jan. 1961, doi: 10.1007/BF02812722.

[17] P. W. Anderson, 'Plasmons, Gauge Invariance, and Mass', *Phys. Rev.*, vol. 130, no. 1, pp. 439–442, Apr. 1963, doi: 10.1103/PhysRev.130.439.

[18] J. Woithe, G. J. Wiener, and F. F. V. der Veken, 'Let's have a coffee with the Standard Model of particle physics!', *Phys. Educ.*, vol. 52, no. 3, p. 034001, Mar. 2017, doi: 10.1088/1361-6552/aa5b25.

[19] P. W. Higgs, 'Broken Symmetries and the Masses of Gauge Bosons', *Phys. Rev. Lett.*, vol. 13, no. 16, pp. 508–509, Oct. 1964, doi: 10.1103/PhysRevLett.13.508.

[20] F. Close, *The Infinity Puzzle: The personalities, politics, and extraordinary science behind the Higgs boson*. Oxford University Press, USA, 2013.

[21] M. Chalmers, 'Higgs@10 – A boson is born', *CERN*, Apr. 28, 2022. https://home.cern/news/news/physics/higgs10-boson-born (accessed May 04, 2022).

[22] J. Ellis, M. K. Gaillard, and D. V. Nanopoulos, 'A phenomenological profile of the Higgs boson', *Nucl. Phys. B*, vol. 106, pp. 292–340, Jan. 1976, doi: 10.1016/0550-3213(76)90382-5.

[23] ALEPH collaboration, DELPHI collaboration, L3 collaboration, OPAL collaboration, and The LEP Working Group for Higgs Boson Searches, 'Search for the Standard Model Higgs boson at LEP', *Phys. Lett. B*, vol. 565, pp. 61–75, Jul. 2003, doi: 10.1016/S0370-2693(03)00614-2.

[24] CERN, 'LEP shuts down after eleven years of forefront research', *CERN*, Nov. 08, 2000. https://home.cern/news/press-release/cern/lep-shuts-down-after-eleven-years-forefront-research (accessed May 23, 2022).

[25] D. A. van Dyk, 'The role of statistics in the discovery of a Higgs boson', *Annu. Rev. Stat. Its Appl.*, vol. 1, pp. 41–59, 2014, doi: 10.1146/annurev-statistics-062713-085841.

[26] C. L. Smith, 'Genesis of the Large Hadron Collider', *Philos. Trans. R. Soc. Math. Phys. Eng. Sci.*, vol. 373, no. 2032, p. 20140037, Jan. 2015, doi: 10.1098/rsta.2014.0037.

[27] CERN, 'Facts and figures about the LHC | CERN', *CERN*. https://home.cern/resources/faqs/facts-and-figures-about-lhc (accessed May 23, 2022).

[28] E. Scott and CMS Collaboration, 'Measurements of the Higgs boson decaying into two photons at CMS', in *Proceedings of The 39th International Conference on High Energy Physics — PoS(ICHEP2018)*, Aug. 2019, vol. 340, p. 260. doi: 10.22323/1.340.0260.

[29] CMS Collaboration and T. Mc Cauley, 'CMS event displays of Higgs to two photon candidate', *CERN Document Server*, Sep. 29, 2020. https://cds.cern.ch/record/2736135 (accessed Apr. 25, 2022).

[30] J. Incandela, F. Gianotti, and R. Heuer, 'Latest update in the search for the Higgs boson', CERN, Geneva, Jul. 04, 2012. Accessed: May 04, 2022. [Online]. Available: https://indico.cern.ch/event/197461/

[31] ATLAS Collaboration and CMS Collaboration, 'Combined Measurement of the Higgs Boson Mass in p p Collisions at s = 7 and 8 TeV with the ATLAS and CMS Experiments', *Phys. Rev. Lett.*, vol. 114, no. 19, p. 191803, May 2015, doi: 10.1103/PhysRevLett.114.191803.

[32] M. Brice, 'Higgs collection images gallery', *CERN Document Server*, Jul. 04, 2012. https://cds.cern.ch/images/CERN-HI-1207136-62 (accessed Apr. 25, 2022).

[33] N. G. Lederman, 'Nature of science: Past, present, and future', in *Handbook of research on science education*, Routledge, 2007, pp. 845–894.

[34] F. Abd-El-Khalick, 'Examining the Sources for our Understandings about Science: Enduring conflations and critical issues in research on nature of science in science education', *Int. J. Sci. Educ.*, vol. 34, no. 3, pp. 353–374, Feb. 2012, doi: 10.1080/09500693.2011.629013.

[35] M. Cápay and M. Magdin, 'Tasks for teaching scientific approach using the black box method', in *European Conference on e-Learning (ECEL)*, 2013, pp. 64–70.

[36] D. Llewellyn, *Inquire within: Implementing inquiry-and argument-based science standards in grades 3-8*. Corwin press, 2013.

[37] S. Miller, 'Modeling the nature of science with the mystery tube', *Phys. Teach.*, vol. 52, no. 9, pp. 548–551, 2014, doi: https://doi.org/10.1119/1.4902200.

[38] M. Krell and S. Hergert, 'The black box approach: analyzing modeling strategies', in *Towards a competence-based view on models and modeling in science education*, Springer, 2019, pp. 147–160.

[39] AG Didaktik der Physik, FU Berlin, 'Modelle in der Biologie', *Tetfolio FU Berlin*. https://tetfolio.fu-berlin.de/web/440484 (accessed Apr. 24, 2022).

[40] N. Lederman and F. Abd-El-Khalick, 'Avoiding de-natured science: Activities that promote understandings of the nature of science', in *The nature of science in science education*, Springer, Dordrecht, 1998, pp. 83–126. [Online]. Available: doi.org/10.1007/0-306-47215-5_5

[41] J. Lederman, S. Bartels, N. Lederman, and D. Gnanakkan, 'Demystifying nature of science', *Sci. Child.*, vol. 52, no. 1, p. 40, 2014, doi: 10.2505/4/sc14_052_01_40.

[42] A. Upmeier zu Belzen and R. Merkel, 'Einsatz von Fällen in der Lehr- und Lernforschung', in *Methoden in der naturwissenschaftsdidaktischen Forschung*, D. Krüger, I. Parchmann, and H. Schecker, Eds. Berlin, Heidelberg: Springer, 2014, pp. 203–212. doi: 10.1007/978-3-642-37827-0_17.

[43] M. Göhner and M. Krell, 'Was ist schwierig am Modellieren? Identifikation und Beschreibung von Hindernissen in Modellierungsprozessen von Lehramtsstudierenden naturwissenschaftlicher Fächer', *Z. Für Didakt. Naturwissenschaften*, vol. 27, no. 1, pp. 155–180, Dec. 2021, doi: 10.1007/s40573-021-00131-4.

[44] CERN S'Cool Lab, '3D-Printable Mystery Box | S'Cool LAB', *CERN*. https://scoollab.web.cern.ch/3d-printable-mystery-box (accessed Apr. 25, 2022).

[45] SCoolLAB, 'Mystery Boxes', *Thingiverse*, Oct. 04, 2018. https://www.thingiverse.com/thing:3136566 (accessed Jun. 13, 2022).

*Appendix 1. Poster template to guide students during the teaching sequence with mystery boxes*

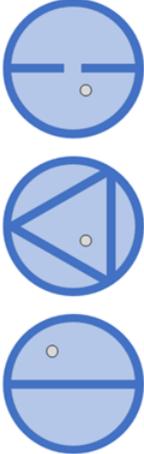